\documentclass[journal]{IEEEtran}

\usepackage{cite}

\usepackage{cite}

\ifCLASSINFOpdf
   \usepackage[pdftex]{graphicx}
   \graphicspath{{../pdf/}{../jpeg/}}
   \DeclareGraphicsExtensions{.pdf,.jpeg,.png}
\else
\fi

\ifCLASSOPTIONcompsoc
 \usepackage[caption=false,font=normalsize,labelfont=sf,textfont=sf]{subfig}
\else
 \usepackage[caption=false,font=footnotesize]{subfig}
\fi

\usepackage{stfloats}

\usepackage{amsmath}                
\usepackage{amsthm}                 
\usepackage{amssymb}

\theoremstyle{definition}
\newtheorem{defn}{Definition}

\usepackage[colorlinks=true, citecolor=blue, urlcolor=blue, linkcolor=blue]{hyperref}       
\usepackage{multicol}
\usepackage{lipsum}

\begin{document}

\title{A Reference Architecture for the American Multi-Modal Energy System}

\author{Dakota~J.~Thompson,~\IEEEmembership{~Thayer School of Engineering at Dartmouth College,}
        Amro~M.~Farid,~\IEEEmembership{~Thayer~School~of~Engineering~at~Dartmouth~College}}


%

\markboth{IEEE Systems Journal (submitted),~Vol.~XX, No.~X, December~2020}%
{Shell \MakeLowercase{\textit{et al.}}: Bare Demo of IEEEtran.cls for IEEE Journals}

\maketitle

\begin{abstract}
The American Multimodal Energy System (AMES) is a system-of-systems comprised of four separate but interdependent infrastructure systems: the electric grid, the natural gas system, the oil system, and the coal system.  Their interdependence creates the need to better understand the underlying architecture in order to pursue a more sustainable, resilient and accessible energy system.  Collectively, these requirements necessitate a sustainable energy transition that constitute a change in the AMES instantiated architecture; although it leaves its reference architecture largely unchanged.  Consequently, from a model-based systems engineering perspective, identifying the underlying reference architecture becomes a high priority.  This paper defines a reference architecture for the AMES and its four component energy infrastructures in a single SysML model.  The architecture includes (allocated) block definition and activity diagrams for each infrastructure.  The reference architecture was developed from the S\&P Global Platts (GIS) Map Data Pro data set and the EIA Annual Energy Outlook dataset.  This reference architecture serves as the foundation from which to accurately and consistently create mathematical model of the AMES.
\end{abstract}

\begin{IEEEkeywords}
American multi-modal energy system, Model-Based Systems Engineering, Reference Architecture, SysML, Energy Systems
\end{IEEEkeywords}

%

\section{Introduction}
\IEEEPARstart{T}{he} American Multimodal Energy System (AMES) is a \textbf{\emph{system-of-systems}} comprised of four separate but interdependent infrastructure systems.  The electric grid, natural gas system, oil system, and coal system comprise the essential infrastructure that meet the energy demands of the 21$^{st}$ century in America.  While each of these individual systems constitute a value chain in their own right, they also enable and support the value chains in the other energy systems.  This interdependence creates the need to better understand the underlying architecture in order to pursue a more sustainable, resilient and accessible energy system.  Each of these three general requirements are discussed in turn.  

\begin{figure*}[t]
\centering
\includegraphics[width=6.5in]{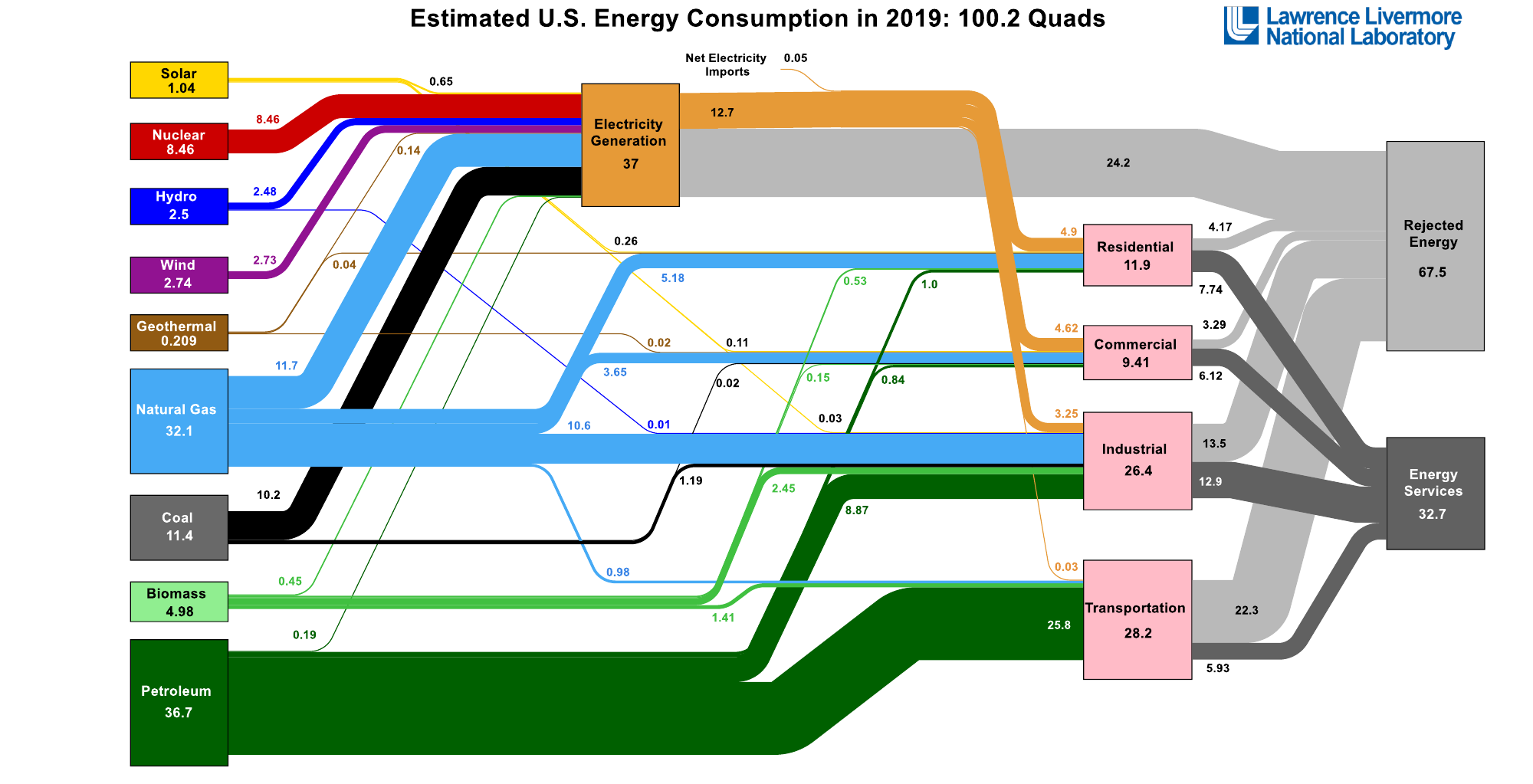}
\caption{A Sankey Diagram of U.S. Energy Consumption in 2019.  The Lawrence Livermore National Laboratory has produced this visualization based primary data sources from the DOE and EIA\cite{LLNL:2019:00}.}
\label{sankey}
\vspace{-0.2in}
\end{figure*}

From a sustainability perspective, the decarbonization of the AMES to meet a global target of not more than a $2^{\circ}$C rise by 2050 is paramount\cite{Elmqvist:2019:00,IEA:2016:00,Birol:2013:00,Commission:2011:00,Rogelj:2016:00,Obergassel:2016:00,williams:2015:00,williams:2015:01,state-of-california:2017:00,IEA:2017:00}.  Graphically, the Sankey diagram developed by the Lawrence Livermore National Laboratory and shown in Fig. \ref{sankey} depicts the AMES flow of energy from primary fuels to four energy consuming sectors\cite{LLNL:2019:00}.  It reveals that the three carbon-intensive fuels of natural gas, petroleum, and coal account for 80\% of the AMES supply side.  In the meantime, 37\% of American energy supply and more importantly  100\% of renewable energy supply flows through electric generation facilities where they are then routed to the residential, commercial, industrial and transportation sectors.  On the demand side, 67\% of all energy consumed is lost as rejected energy.  The transportation sector, in particular, rejects $~80\%$ of its energy and is consequently the lead producer of greenhouse gas (GHG) emissions\cite{EIA:2020:00}.  To significantly reduce the GHG emissions produced from fossil fuels, three \textbf{\emph{architectural changes}} are simultaneously required\cite{IEA:2016:00}. First, carbon-neutral renewable energy sources such as solar, wind, nuclear, geothermal and nuclear generation must be increasingly integrated into the grid and ultimately displace fossil-fuel fired generation plants; especially as they are retired at the end of their useful life\cite{Kamphuis:2008:00,hansen:2001:00,kumar:2016:00,IEA:2017:00,Munoz-Hernandez:2013:00,Chaabene:1998:00}.  Second, energy consumption technologies, like transportation and heating, that rely heavily on fossil-fuel combustion must switch fuels to electricity where they have opportunity to be powered by an increasingly decarbonized electric power.  Lastly, energy-intensive technologies throughout the AMES must be systematically replaced with their more energy-efficient counterparts\cite{Pasaoglu:2012:00,Litman:2013:00,Andersen:2009:00,Sortomme:2012:00,U.S.-Energy-Information-Administration:2015:00,Anair:2012:00}.  

Together, these three architectural changes minimize the demand on the coal, oil, and natural gas systems.  In the meantime, such a systemic shift towards the use of electricity requires a commensurate expansion of the electric grid.  Such a \textbf{\emph{sustainable energy transition}} is arguably the largest single engineering system transformation in human history.  Given the environmental consequences, the energy transition must be undertaken in a manner that not just meets the evolving requirements of its stakeholders, but also remains operational.  Fortunately, from a Model-Based Systems Engineering (MBSE) perspective\cite{Dori:2015:00,Friedenthal:2011:00}, the three architectural changes described above constitute a change in the AMES \emph{instantiated architecture} but leaves the AMES \emph{reference architecture} largely unchanged.  In order to deploy an MBSE-methodology to the sustainable energy transition, identifying the underlying reference architecture of the AMES becomes a high priority in meeting the paramount requirement of energy sustainability.  

\begin{defn}\textbf{- Instantiated Architecture}\cite{Cloutier:2010:00,Friedenthal:2011:00} A case specific architecture, which represents a real-world scenario, or an example test case. At this level, the physical architecture consists of a set of instantiated resources, and the functional architecture consists of a set of instantiated system processes. The mapping defines which resources perform what processes.
\end{defn}

\begin{defn}\textbf{- Reference Architecture\cite{Cloutier:2010:00}} \textit{``The reference architecture captures the essence of existing architectures, and the vision of future needs and evolution to provide guidance to assist in developing new instantiated system architectures. ...Such reference architecture facilitates a shared understanding across multiple products, organizations, or disciplines about the current architecture and the vision on the future direction. A reference architecture is based on concepts proven in practice. Most often preceding architectures are mined for these proven concepts. For architecture renovation and innovation validation and proof can be based on reference implementations and prototyping. In conclusion, the reference architecture generalizes instantiated system architectures to define an architecture that is generally applicable in a discipline. The reference architecture does however not generalize beyond its discipline."}
\end{defn}

The primary benefit of a reference architecture is that it clearly identifies the system boundary, the components of the system form, the activities of the system behavior, and the interfaces and interactions between them.  This identification is of critical importance when the chosen system is particularly complex and heterogeneous; as in the case of the AMES.  In the electric power system (alone), there is a rich history of reference architecture development in the so-called ``Common Information Model"\cite{Uslar:2012:00,Gray:2019:00} that has culminated in IEC Standards 61970, 61968, and 62325\cite{IEC:2012:00,IEC:2013:00,IEC:2014:00}.  Furthermore, it is important to recognize that a reference architecture, by design, can admit a wide variety of mathematical models of system behavior.  For example, once the relevant classes of an electric power system have been identified in a reference architecture, depending on the need, one can still develop an AC or DC power flow analysis model, an AC or DC optimal power flow model, or a small signal or transient stability model.  Naturally, the choice of mathematical modeling elements that are being superimposed on the reference architecture greatly affects the computational intensity of the mathematical model as a whole.  Additionally, the chosen mathematical model may be implemented as a computational (simulation) model that is either centralized (on one processor) or distributed (on many).  Furthermore, depending on the causal dependencies in the reference architecture, a distributed computational model may invoke fully parallel processing, or sequential co-simulation techniques\cite{godfrey:2010:00,gomes:2018:00}.  This work leaves these mathematical modeling and computational implementation as choices outside the scope of this paper, but ultimately recognizes that the development of a reference architecture is a necessary first step.

From a resilience and reliability perspective, each of the AMES component systems must not just deliver their respective type of energy independently\cite{Elmqvist:2019:00,Rueda:2017:00,Uday:2015:00,Thompson:2020:SPG-JR05,chassin:2005:00} but must also support the other AMES infrastructures\cite{Hernandez-Fajardo:2013:00}.  For example, and as shown in Fig. \ref{sankey}, if a natural gas pipeline fails in the natural gas system it could take a natural gas power plant offline in the electric grid.  Such a lack of electric generation capacity could then result in the temporary shut down of a natural gas processing plant; further reducing natural gas and electricity capacity.  The New England electric power grid, in particular, remains susceptible to natural gas shortages during long cold spells when the fuel is used heavily for both space heating as well as electric generation\cite{ICF:2012:00}.  Alternatively, the oil and natural gas systems rely on electricity to process their respective fuels and compress them during storage and transportation.  Even the coal system requires electricity in safe and efficient mining.  

As the AMES architecture evolves through the sustainable energy transition, it must do so in a manner that is reliable and resilient to natural, economic and malicious disruptions.  By modeling and understanding the instantiated architecture of the AMES at each stage of this sustainable energy transition, system-wide vulnerabilities can be systematically identified and mitigated in a way that is more comprehensive than if each infrastructure were studied independently. For example, global climate change and severe weather events may place coastal energy facilities particularly at risk\cite{Bonham:2020:00}.  Additionally, economic shocks can affect the import and export energy resources and disrupt their relative balance in the AMES\cite{venn:2016:00}.  Finally, malicious cyber-attacks can propagate failures not just within a given AMES infrastructure but across them as well.  

Finally, from an energy access perspective, the AMES must continue to cost-effectively and equitably provide readily available energy to the broader public \cite{action:2016:00}.  Relative to many other nations, this requirement has been largely addressed in the United States.  Nevertheless, certain issues remain.  For example, in northern New England, people rely on carbon-intensive oil and propane for heating since heat pumps have limited performance in especially cold climates.  Finally, solar and wind potential is often plentiful away from urban load centers and so may not be effectively tapped without additional electric transmission capacity\cite{Gellings:2011:00,Gungor:2011:00,Gungor:2013:00,iso-C:2012:00,Joos:2000:00,Xie:2011:00,Pickard:2009:00,Kassakian:2011:00}.  Many of these energy access concerns are particularly poignant in Alaska and other arctic regions.    

The three general requirements of energy sustainability, resilience, and access impose constraints on the evolution of the AMES architecture.  And yet, the AMES architecture remains relatively poorly understood from a holistic perspective\cite{pietzcker:2017:00,Haller:2012:00,howard:2014:00,Rogers:2014:00}.  The Sankey Diagram in Fig. \ref{sankey}, to our knowledge, presents the only graphical depiction of the AMES in its entirety.  While this data visualization effectively conveys information concerning relative energy flows, from a model-based systems engineering\cite{Dori:2015:00} perspective, its highly simplified nature was not intended for architectural analysis and design.  In addition to the Sankey model, the EIA has developed the National Energy Modeling System (NEMS) software to produce the yearly annual energy outlook\cite{EIA:2020:00}.  Nevertheless, this software-based tool remains less than transparent and the EIA website itself states:  ``[The] NEMS is only used by a few organizations outside of the EIA. Most people who have requested NEMS in the past have found out that it was too difficult or rigid to use \cite{EIA:2017:01}". 

\begin{figure*}[!t]
\centering
\includegraphics[width=6.5in]{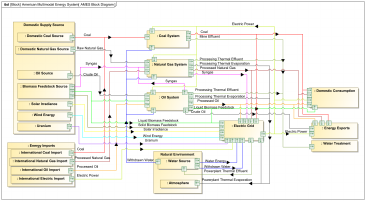}
\caption{The top level internal block diagram of the AMES.  The domestic supply sources, the energy imports, natural environment, domestic consumption, energy exports, and water treatment are external to the AMES four subsystems of coal, natural gas, oil, and electric grid.  }
\label{AMESModel}
\vspace{-0.2in}
\end{figure*}

\vspace{-0.15in}
\subsection{Original Contribution}
In order to deploy an MBSE-methodology to the sustainable energy transition, this paper uses a \textbf{\emph{data-driven approach}} to define a reference architecture in a single invariant SysML model describing the four main systems that comprise the unified AMES.  By defining the reference architecture, this paper provides the foundation from which to consistently build an instantiated architecture for future mathematical modeling. The top level block diagram in Fig. \ref{AMESModel} presents the four subsystems of the AMES and the flow of operands between them and those entities defined as outside of the system boundary.  Each of the four subsystems: electric grid, natural gas system, oil system, and coal system are in turn defined using class and activity diagrams with (allocation) swim-lanes.  Integrating each of the sub-reference architecture class and activity diagrams as described in the AMES block diagram defines the entirety of the AMES reference architecture.  This work assumes a working knowledge of the SysML (the Systems Modeling Language) which is otherwise gained from several excellent texts\cite{Dori:2015:00,Friedenthal:2011:00,Weilkiens:2007:00}.  

\vspace{-0.15in}
\subsection{Paper Outline}
Section \ref{background} starts with a description of the background literature and the datasets used to develop the reference architecture.  The paper then presents the electric power system's architecture in Section \ref{model_elec}.  The natural gas architecture is then presented in Section \ref{model_NG}.  The oil system and coal system architectures are then defined in Section \ref{model_oil} and Section \ref{model_coal} respectively.  A discussion of dependencies between each of the subsystems is presented in Section \ref{discussion}.  The paper then presents future work of the AMES reference architecture.  This includes simulation development for integrated planning and operations management.  Finally, the paper is brought to a conclusion in Section \ref{conclusion}.



\vspace{-0.1in}
\section{Background}
\label{background}
Normally, each of the four systems of the AMES are studied independently and each have their own extensive literature\cite{masters:2013:00,mokhatab:2012:00,Lurie:2009:00,EIA:2014:11}.  Increasingly, however, sustainability, resilience, and accessibility drivers have brought about greater attention to how these individual infrastructures depend on each other\cite{Albert:2004:00,Dong:2015:00,Jean-Baptiste:2003:00,Kriegler:2018:00,Mejia-Giraldo:2012:00,Robert-Lempert:2019:00,Rogers:2013:00}.  One dependence that has received considerable attention is the dependence of the electric grid on the natural gas system \cite{Al-Douri:2017:00,An:2003:00,ICF:2012:00,Li:2008:01,Shahidehpour:2005:01,Unsihuay:2007:00,Zlotnik:2017:00,Jenkins:2015:00}.  These works are motivated by the increasing role of natural gas-fired electricity generation relative to coal-fired facilities\cite{kerr:2010:00}, and the importance of natural gas power plants in providing ``flexible" operating reserves against variable renewable energy resources\cite{Muzhikyan:2019:SPG-J39}.  Similarly, some works have addressed the dependence of the electric grid on the oil\cite{Lurie:2009:00,aleklett:2010:00} and coal systems \cite{AAR:2016:00,EIA:2014:11,Mejia-Giraldo:2012:00}.  Moving beyond the specific scope of the AMES, a related but extensive literature has developed on the co-dependence of the electric grid and water resources in the form of the Energy Water Nexus (EWN) \cite{Farid:2016:EWN-J29,Lubega:2014:EWN-J11,Lubega:2014:EWN-J12,Thompson:2019:00,Munoz-Hernandez:2013:00,Lubega:2014:EWN-T09,Lubega:2013:EWN-C36,Lubega:2014:EWN-C35,Lubega:2014:EWN-C34,Farid:2013:EWN-C15,Santhosh:2012:EWN-C09,Santhosh:2013:EWN-C26,Santhosh:2014:EWN-J10,Hickman:2017:EWN-J32,Murrant:2015:00,Wakeel:2016:00,Macknick:2012:00,Macknick:2012:01,Averyt:2011:00,Tidwell:2014:00}.  Together, these works provide an insight into the structural and behavioral complexity of the AMES.  Furthermore, they also demonstrate the potential benefits of analyzing and optimizing the AMES as a single system-of-systems rather than each system independently\cite{Roosa:2008:00}.  Other works have sought to model multi-energy systems\cite{mancarella:2016:00,subramanian:2018:00,Beuzekom:2015:00,Geidl:2007:00,Krause:2011:00,Ma:2018:00,MANCARELLA:2014:00,Quelhas:2007:00,WU:2016:00,jacobson:2009:00,jacobson:2009:01,jacobson:2015:00,jenkins:2017:00,jenkins:2018:00} making use of energy hubs to facilitate and track the flows of energy often focusing on the interactions of electricity, natural gas, distributed heating, and renewable sources.  These approaches lack a SysML approach to explicitly define each component with their allocated functions and associations within the defined architecture.

It is worth mentioning that much of these works focus on a single interaction between two energy systems and consequently, to our knowledge, this is the first work to address the architecture of the AMES as a whole modeling electric, natural gas, oil, and coal systems  in a single invariant SysML model.  Furthermore, because the focus is usually on a single interaction, there has been little effort\cite{Lubega:2014:EWN-J11,Lubega:2013:EWN-C19,Abdulla:2015:EWN-C53} to deploy a model-based systems engineering methodology where a system boundary is rigorously defined and then later elaborated in terms of physical interfaces and functional interactions.  Ultimately, a complete architectural description is necessary to ensure that 1.) energy and mass conservation laws are respected, 2.) all environmental aspects are identified in environmental impact assessments\cite{glasson:2013:00}, and 3.) the greatest potential for synergistic outcomes are found.  Finally, the use of model-based systems engineering modeling conventions (such as SysML) maximizes the potential for cross-disciplinary communication and coordination.

This paper takes a \textbf{\emph{data-driven approach}} and uses the S\&P Global Platts (GIS) Map Data Pro data set\cite{Platts:2017:00} and the EIA Annual Energy Outlook dataset \cite{EIA:2020:00} to deduce the AMES reference architecture.  The S\&P Global Platts (GIS) Map Data Pro data set\cite{Platts:2017:00} is a proprietary data set available through the S\&P Global Platts website.  It is labeled with metadata that correspond to classes and attributes in the AMES form.  The classes and their associated behaviors are shown here, but their attributes have been suppressed for brevity.  The interested reader is referred to original references for attribute metadata.  Next, each GIS layer of the Platts dataset includes descriptions of facility types and their associated products.  This data can be used to deduce the associated function(s) of these facilities.  Finally, the process technologies for all of the AMES constituent energy facilities are well known.  Therefore, this work relies on engineering textbook knowledge of these facilities to supplement the Platts and EIA datasets with low-level knowledge of input-output interfaces.  

While the choice of a \textbf{\emph{data-driven approach}} leads straightforwardly to a well-validated reference architecture model, it is not without its limitations.  First, the scope of this work is limited to only the energy systems themselves and not the end-use sectors outside of the AMES.  Second, because Platts and EIA datasets only include bulk, wholesale, and transmission level assets,  distribution-level and retail-level assets are outside the scope of the work.  Finally, any assets outside of the conventional electric grid, natural gas system, oil system, and coal system are naturally out of scope as well.  This includes non-conventional energy technologies and carriers (e.g. bio-energy, hydrogen, ammonia, etc) which have yet to make a sizable impact on American energy infrastructure. 

\vspace{-0.1in}
\section{Modeling}\label{modeling}

This paper uses the Systems Modeling Language (SysML)\cite{Friedenthal:2011:00,Reichwein:2012:00,SE-Handbook-Working-Group:2011:00,Weilkiens:2007:00, Crawley:2015:00,Dori:2015:00,Rumbaugh:2005:00} to define the AMES reference architecture.  More specifically, the metadata of the input datasets are conserved, reorganized and drawn within SySML block definition and activity diagrams.  This data-driven approach produces a SysML reference architecture that includes: 1.) the different facilities that comprise the AMES form and 2.) their processes that comprise the AMES functionality and allocated architecture. Fig. \ref{AMESModel} shows the system boundary of the AMES around its four constituent energy systems of electricity, oil, natural gas and coal.  The high level flows of matter and energy between these four energy systems and across the system boundary are also defined. The matter and energy flows in Fig. \ref{AMESModel} also restrict the set of operands in the AMES.  While the Platts dataset does specify a much larger number of energy products, this analysis, for tractability, has classified all flows of matter and energy into the following set of operands:   coal, raw natural gas, processed natural gas, crude oil, processed oil, syngas, liquid biomass feedstock, solid biomass feedstock, solar irradiance, wind energy, uranium, water energy (for cooling), electric power, withdrawn water, mine effluent, processing effluent, and thermal effluent.  Therefore, Fig. \ref{AMESModel} shows the input flow of these quantities of matter/energy operands from the domestic supply sources, the energy imports, and the natural environment across the system boundary and the output flow of these quantities to domestic consumption, energy exports, water treatment facilities, and the natural environment.  In all cases, these input/output flows are specified in mass flow rates (e.g. Kg/time) or power (W) or both where the associated matter has an intrinsic energy content (e.g. heating value of natural gas).  

\begin{figure}[!h]
\vspace{-0.1in}
\centering
\includegraphics[width=3.45in]{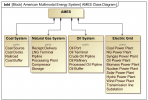}
\vspace{-0.3in}
\caption{AMES block definition diagram showing its four component systems.}
\label{AMESClass}
\vspace{-0.1in}
\end{figure}

From a form perspective, Fig. \ref{AMESClass} presents a class diagram of the AMES and its four constituent energy systems as classes.   For graphical simplicity, each of these energy system classes adopt attributes to represent their component infrastructure facilities and resources.  Furthermore, association links are removed for graphical clarity and may be otherwise deduced from the associated activity diagram.  The following subsections elaborate the form and function of these systems.  

\vspace{-0.1in}
\subsection{Electric Power System}
\label{model_elec}
\begin{figure}[!h]
\centering
\includegraphics[width=3.45in]{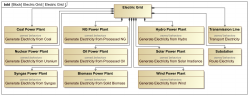}
\caption{Electric grid block definition diagram showing its component physical resources.}
\label{ElecClass}
\vspace{-0.1in}
\end{figure}

The Electric Power System is comprised of resources for the generation, transmission, and routing of electric power.  Power plants comprise a majority of the different types of resources within the electric grid.  Each power plant type is designated by the primary fuel category used to generate electric power.  There are nine different types of power plants present: coal, natural gas, syngas, oil, biomass, nuclear, solar, hydro, and wind.  These power plants are connected to the electric grid by transmission lines (to the distribution system).  The last component of the electric grid that realizes the end of the electric grid value chain is substations where the electric power leaves the transmission system. Fig. \ref{ElecClass} presents the formal decomposition of the AMES electric grid architecture.

\begin{figure*}[!t]
\centering
\includegraphics[width=6.5in]{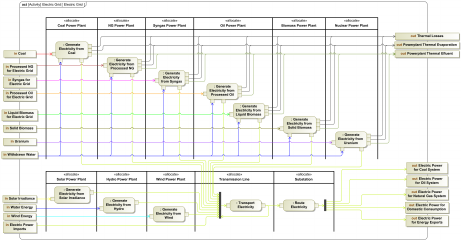}
\caption{Electric grid activity diagram with allocated swim-lanes.}
\label{ElecAct}
\vspace{-0.2in}
\end{figure*}

Each of the individual resources within the electric power system have their respective processes.  Fig. \ref{ElecAct} presents the electric grid activity diagram that shows these processes allocated onto their respective form in swim-lanes and follows the flows of matter and energy between the processes.  Each power plant has their respective generate electric power process from their designated fuel source.  The thermal generation processes Generate Electricity from Coal, Generate Electricity from Processed NG, Generate Electricity from Syngas, Generate Electricity from Processed Oil, Generate Electricity from Liquid Biomass, Generate Electricity from Solid Biomass, and Generate Electricity from Uranium each take their respective fuel source and withdrawn water as inputs and result in electric power, thermal losses, power plant thermal effluent, and power plant thermal evaporation.  Aside from electric power, all of the remaining outputs immediately leave the system boundary.  In contrast, the electric power is then transported by the transmission lines.  The electric grid value chain is completed at the substation which routes the electric power to the other AMES energy systems or to the electric distribution system outside the scope of this reference architecture.

\vspace{-0.1in}
\subsection{Natural Gas System}
\label{model_NG}

\begin{figure}[!htb]
\vspace{-0.1in}
\centering
\includegraphics[width=3.45in]{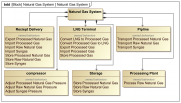}
\vspace{-0.25in}
\caption{Natural Gas system block definition diagram showing its component physical resources.}
\label{NGClass}
\end{figure}

The natural gas system is comprised of resources for the import, export, processing and delivery of natural gas.  The receipt delivery and Liquefied Natural Gas (LNG) terminals are responsible for importing and exporting natural gas into and out of the natural gas system.  These resources take both international and domestic imports into the United States' natural gas pipeline infrastructure.  Pipelines and compressors are present for facilitating the transportation of natural gas.  Additionally, processing plants are present for processing raw natural gas.  Finally, storage facilities store syngas as well as raw and processed natural gas.  Fig. \ref{NGClass} presents the formal decomposition of the AMES natural gas system architecture.

\begin{figure*}[!tb]
\centering
\includegraphics[width=6.5in]{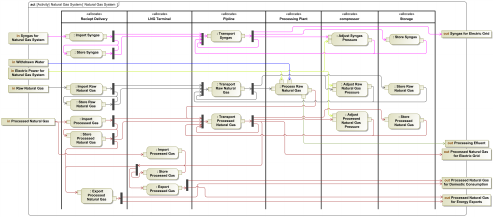}
\vspace{-0.1in}
\caption{Natural gas system activity diagram with allocated swim-lanes.}
\label{NGAct}
\end{figure*}

Each of the individual resources within the natural gas system have their respective processes.  Fig. \ref{NGAct} presents the natural gas activity diagram.  It shows natural gas processes allocated onto their respective form in swim-lanes and follows their flow of matter and energy.  The Receipt Delivery facility can import and store syngas, raw natural gas, and processed natural gas as well as export the processed natural gas out of the system boundary.  The LNG Terminal can import, store and export natural gas.  Once inside the natural gas system, pipelines transport each of the operands, syngas, raw natural gas and processed natural gas, through the United States.  This includes pipelines that transport directly to natural gas electric power plants in the electric grid.  With the inputs of raw natural gas, electric power and withdrawn water, processing plants process raw natural gas to produce processed natural gas and processing effluent.  Compressors stimulate the transportation of the different types of natural gas by adjusting the associated pressure.   Finally, storage facilities store syngas as well as raw and processed natural gas.  

\vspace{-0.1in}
\subsection{Oil System}\label{model_oil}

\begin{figure}[!h]
\vspace{-0.1in}
\centering
\includegraphics[width=3.45in]{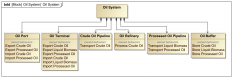}
\vspace{-0.25in}
\caption{Oil system block definition diagram showing its component physical resources.}
\label{OilClass}
\vspace{-0.1in}
\end{figure}

The oil system is comprised of resources for the import, export, and delivery of oil.  The oil port and oil terminal are responsible for importing and exporting oil into and out of the oil system.  These resources take both international and domestic imports into the United States' oil pipeline infrastructure.  Crude and processed oil pipelines are present for facilitating the transportation of oil and liquid biomass.  Oil refineries allow for the processing of crude oil into processed oil, and oil buffers allow for storage within the oil system infrastructure.  Fig. \ref{OilClass} presents the formal decomposition of the AMES oil system architecture.

\begin{figure*}[!htb]
\centering
\includegraphics[width=6.5in]{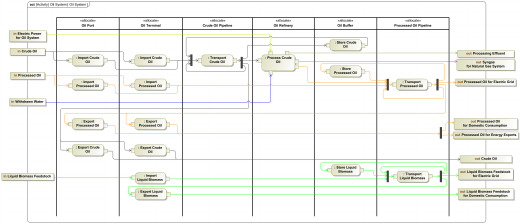}
\vspace{-0.1in}
\caption{Oil system activity diagram with allocated swim-lanes.}
\label{OilAct}
\end{figure*}

Each of the individual resources within the oil system have their respective processes.  Fig. \ref{OilAct} presents the oil activity diagram.  It shows the oil system's processes allocated onto their respective form in swim-lanes and follows their flows of matter and energy.  The Oil Terminal facility can import and export crude oil, processed oil and liquid  biomass to and from outside the system boundary.  The Oil Port can also import and export crude and processed oil.  Once inside the oil system, the crude oil pipeline can transport crude oil from an oil port or terminal to an oil refinery where the crude oil is processed into processed oil.  This process requires the input of crude oil, electricity and withdrawn water to produce processed oil, syngas\cite{rosa:2017:00} and processing effluent.  The processed oil can then be transported by the processed oil pipelines. These processed oil pipelines transport processed oil and liquid biomass within the oil system and directly to oil and liquid biomass electric power plants in the electric grid.  Additionally, all three operands, crude oil, processed oil, and syngas can be stored within the oil system by oil buffers.  

\vspace{-0.15in}
\subsection{Coal System}
\label{model_coal}

\begin{figure}[!htb]
\centering
\includegraphics[width=3.2in]{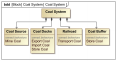}
\vspace{-0.1in}
\caption{Coal system block definition diagram showing its component physical resources.}
\vspace{-0.2in}
\label{CoalClass}
\end{figure}

The coal system is comprised of resources for the import, export, and delivery of coal.  The coal sources are responsible for mining domestic sources of coal and introducing coal into the United States coal system.  Coal docks are also responsible for the import and export of coal.   Railroads are responsible for transporting coal across the United States and to coal electric power plants in the electric grid.  Finally, coal buffers allow for the storage of coal within the system boundary. Fig. \ref{CoalClass} presents the formal decomposition of the AMES coal system architecture.

\begin{figure*}[!htb]
\centering
\includegraphics[width=6.5in]{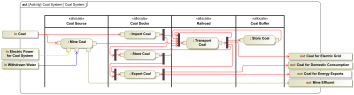}
\vspace{-0.1in}
\caption{Coal system activity diagram with allocated swim-lanes.}
\label{CoalAct}
\vspace{-0.2in}
\end{figure*}

Each of the individual resources within the coal system have their respective processes.  Fig. \ref{CoalAct} presents the coal activity diagram.  It shows these processes allocated onto their respective form in swim-lanes and follows their flow of matter and energy.  With the input of electric power and withdrawn water, the coal source can mine coal to produce coal and mine effluent.  Alternatively, the coal docks can import coal into the coal system which can then be transported by the railroads.  The coal can then be stored within a coal buffer or exported out of the coal system boundary by a coal dock.

\vspace{-0.1in}
\section{Discussion}\label{discussion}

The activity diagrams in Figs. \ref{ElecAct}, \ref{NGAct}, \ref{OilAct}, \ref{CoalAct} each show the individual energy systems that when integrated together form the AMES shown in Fig. \ref{AMESModel}.  When following the flows of matter and energy through the AMES, it becomes apparent that every subsystem is connected to the other.  The coal system produces and imports coal that is delivered to the electric grid for electric power generation.  The oil system is able to produce and deliver syngas to the natural gas system as well as deliver processed oil and liquid biomass to the electric grid for electric power generation.  The natural gas system is able to deliver syngas and processed natural gas to the electric grid for electric power generation.  Finally, the electric grid is able to deliver electric power to the coal system for mining, the oil system for processing crude oil, and to the natural gas system for processing raw natural gas and gas compression.  Each of these connections allow the electric grid to produce electric power from the other energy systems' fuel sources and subsequently deliver power to the United States.  Additionally, the electric power allows for the production and processing of operands in the coal, oil, and natural gas systems so that they may provide fuel sources back to the electric grid and the rest of the United States' fuel demands.  

Understanding the nature of such interdependencies within the AMES reference architecture facilitates changes to the AMES as it is currently instantiated\cite{De-Weck:2011:00,Rinaldi:2004:00,Prasad:2014:00}.  This knowledge becomes particularly important in avoiding cross-sectoral cascading failures \cite{Dong:2015:00,buldyrev:2010:00,Uday:2015:00}.  For example, if a natural gas pipeline fails, there is not only a loss of natural gas being delivered for heating, but for electric power generation as well.  Unavailable electric power plants not only diminishes the grid's ability to meet residential, commercial, and industrial demand, but also the load demanded by the other energy systems.  

These interdependencies in the AMES reference architecture often exaggerate ``infrastructure lock-in" effects that impede the forward-motion of the sustainable energy transition\cite{unruh:2000:00,unruh:2002:00,seto:2016:00,markolf:2018:00}.  As coal power plants are decommissioned, natural gas power plants are often installed in their place with commensurate reductions in greenhouse gas emissions.  These benefits, however, are not realized until sufficient natural gas pipeline capacity is secured; either on existing or potentially new pipelines.   Similarly, electric power transmission capacity often impedes the full utilization of remote solar and wind generation resources.  Alternatively, the presence of excess processing and transmission capacity for coal, oil, and natural gas makes it very easy and economical to rely on these sources in the electric power sector.  For example, the electric power grid is likely to retain its reliance on the natural gas system for a long time because so much of the country relies on natural gas for heating.   In short, an effective ``deep" decarbonization strategy requires the coordination of all four energy sectors and not just one alone.   

\vspace{-0.1in}
\subsection{Integrated Planning and Operations and Model Development}\label{planning_oper}
By planning future infrastructure developments with an integrated view of the whole AMES, developments with the greatest impact can be planned and installed.  This allows for a holistic planning effort that incentivizes simultaneous developments in multiple energy systems such that they compliment, rather than impede, each other.  For example, if a coal mine is decommissioned in the coal system, then a coal power plant in the electric grid could be replaced with a less carbon-intensive power plant.  The EWN literature has already demonstrated similar benefits\cite{Palensky:2011:00,Santhosh:2012:EWN-C09,Santhosh:2013:EWN-C26,Santhosh:2014:EWN-J10,Hickman:2017:EWN-J32,Xie:2009:01}.  For example, the straightforward installation of water storage capacity has been shown to alleviate power balance constraints in the electric power grid where the installation of battery energy storage is at a premium.   Similarly, the natural gas-electricity literature has shown pairing natural gas electric power plants with variable energy resources (VER) such as wind turbines provides a smaller carbon footprint with renewable wind energy and natural gas replacing coal\cite{alabdulwahab:2015:00,Li:2008:01,MACKINNON:2018:00}.  Additionally, the fast ramping capacity of natural gas power plants provides reliability in maintaining a stable grid in the presence of VERs.  In all of these cases, one or more layers of planning and operations management decision-making are superimposed on the underlying interdependent infrastructure system's instantiated mathematical model.   

\vspace{-0.1in}
\subsection{Dynamic Simulation Model Development}
\label{planning_operation}

The development of the AMES reference architecture facilitates the subsequent development of instantiated mathematical models and simulations of system behavior.  As a relevant precedent, the energy-water nexus reference architecture\cite{Lubega:2014:EWN-J11,Lubega:2013:EWN-C19} led to the development of holistic mathematical models\cite{Lubega:2014:EWN-J12,Lubega:2013:EWN-BC02,Lubega:2014:EWN-T09,Lubega:2014:EWN-C34,Farid:2016:EWN-J29} which were later implemented as numerical simulations. 
To this end, the reference architecture provides the starting point for a \textbf{\emph{transparent}} objected-oriented software design grounded in ``digital twin" principles.  Much like the National Energy Modeling System (NEMS)\cite{EIA:2019:00}, the AMES reference architecture can be used to model and simulate the effect of potential policies and future infrastructure developments.  By using the reference architecture's defined components, one can instantiate an existing or simulated architecture.  For example, using the Platts Map Data Pro \cite{Platts:2017:00} to guide the installation of resources and their subsequent associations, a SysML-compliant model of the AMES can be created.  By then varying the ratios of instantiated technologies belonging to the instantiated model variant, different scenarios can be analyzed to further advance the AMES development. 

Just as the AMES reference architecture allows for the simulation and analysis of differing policies across the entirety of the AMES, it also allows for integrated operations management and power flow analysis.  As seen in past energy-water nexus works, the mathematical models were later used to conduct sensitivity analyses and identify input/output trade-offs\cite{Hickman:2017:EWN-J32,Lubega:2014:EWN-C35,Lubega:2014:EWN-J12,Farid:2013:EWN-C15,Abdulla:2015:EWN-C53}.    Such an approach of translating a reference architecture into a dynamic simulation/optimization model has been further generalized using hetero-functional graph theory\cite{Schoonenberg:2018:ISC-BK04}.  As demonstrated in previous work \cite{schoonenberg:2021:00}, the use of continuous, timed, and arc-constant colored Petri nets facilitate the transition of a reference architecture into a (generic) hetero-functional minimum cost flow optimization problem. By introducing a device model for each capability independently, the modeler can control the relevant time scale, and thereby eliminate fast dynamics with steady state approximations.  This allows for each device model to be integrated into the model with the same choice of relevant temporal resolution.  Applying these device models can be viewed as a generalization of a similar procedure that has been demonstrated for electric power system modeling and simulation \cite{Milano:2010:17}.  As each individual capability receives their own device model, the energy flows of differing types throughout the system can all be optimized together towards a well defined objective.  

\vspace{-0.1in}
\subsection{Structural Analysis Model Development}\label{sim_deployment}
Additionally, recent theoretical works\cite{Schoonenberg:2018:ISC-BK04,Thompson:2020:SPG-C68,Thompson:2020:SPG-JR05} have shown that SysML-based reference architectures of interdependent infrastructure systems can be translated, without loss, into mathematical structural models called hetero-functional graphs (HFG).  These HFGs can then be used to study the AMES structural resilience under varying scenarios \cite{Thompson:2020:SPG-C68,Thompson:2020:SPG-JR05}.  By allocating function onto instantiated individual resources, capabilities are formed which can be chained together to complete a value chain and define deliverable services.  Through tracking the number of deliverable services present in the instantiated model, the structural resilience can be analyzed \cite{Thompson:2020:SPG-C68}.  By changing the ratios of these instantiated technologies from the reference architecture, the number of deliverable services will also vary; thereby changing the associated structural resilience.  Alternatively, the instantiated model could be placed under attack, which would dictate the gradual removal of capabilities, and decrease the structural resilience as the instantiated model degrades.  Using the AMES reference architecture to guide the instantiation of various scenarios, future AMES technology compositions can be analyzed to guide the energy transition.

  
\subsection{Pathways to a Sustainable Energy Transition}
Returning to the original motivation of the paper, the AMES reference architecture serves as a critical step in a Model-Based Systems Engineering methodology to the sustainable energy transition.  In that regard, the AMES reference architecture remains invariant while the instantiated architecture undergoes three architectural changes.  The well-received United States Deep Decarbonization Pathways report\cite{williams:2015:00} identifies these three changes as:  1.) the increased penetration of renewable energy technologies into the electric power system 2.) the increased penetration of energy-efficient (consumption) technologies of all types, and 3.) a systematic electrification of energy consuming technologies.  Furthermore, it identified four viable scenarios that mix the relative importance of these three architectural changes.  Because all of these architectural changes are reflected already in the S\&P Global Platts (GIS) Map Data Pro data set, the AMES reference architecture can be instantiated straightforwardly to model these four viable scenarios.  In that regard, the implicit assumption of stationarity in the dataset does not impede modeling and analysis of the AMES instantiated architecture despite the profound changes required by the sustainable energy transition.  

\vspace{-0.1in}
\section{Conclusion}
\label{conclusion}

The American Multi-modal Energy System reference architecture is an invariant reference architecture that describes the electric grid, oil system, natural gas system, and coal system  as well as their  inter-dependencies.  As American energy demands in the 21$^{st}$ evolve to meet new requirements for energy sustainability, resilience, and access, the AMES instantiated architecture will also evolve, but the AMES reference architecture will remain largely unchanged.  Instead, the ratios of instantiated elements will change resulting in more carbon-intense resources being instantiated less and carbon-lite or carbon-free resources being instantiated more.  This AMES reference architecture provides the basis from which to run simulations on new policies and the associated changes of instantiated architecture.  Furthermore, the AMES reference architecture facilitates the formulation of new optimal planning and operations management decisions.   As previously demonstrated in the NG-Electricity nexus literature and the energy-water nexus literature, these decisions can identify synergistic strategies that simultaneously enhance infrastructure cost, reliability and sustainability.   Such synergistic strategies are often able to overcome typical ``infrastructure lock-in" scenarios and the ensuing ``trilemma" debates on energy sustainability, resilience, and access.  In short, holistic AMES models present new possibilities for energy infrastructure coordination that may have been otherwise overlooked when addressing each energy infrastructure independently.  Through future work exploring the static and dynamic simulations of the AMES, this reference architecture provides the first step towards guiding the energy transition.

\vspace{-0.1in}

\bibliographystyle{IEEEtran}
\bibliography{AMESArchitectureLib.bib}

\begin{thebibliography}{100}
\providecommand{\url}[1]{#1}
\csname url@samestyle\endcsname
\providecommand{\newblock}{\relax}
\providecommand{\bibinfo}[2]{#2}
\providecommand{\BIBentrySTDinterwordspacing}{\spaceskip=0pt\relax}
\providecommand{\BIBentryALTinterwordstretchfactor}{4}
\providecommand{\BIBentryALTinterwordspacing}{\spaceskip=\fontdimen2\font plus
\BIBentryALTinterwordstretchfactor\fontdimen3\font minus
  \fontdimen4\font\relax}
\providecommand{\BIBforeignlanguage}[2]{{%
\expandafter\ifx\csname l@#1\endcsname\relax
\typeout{** WARNING: IEEEtran.bst: No hyphenation pattern has been}%
\typeout{** loaded for the language `#1'. Using the pattern for}%
\typeout{** the default language instead.}%
\else
\language=\csname l@#1\endcsname
\fi
#2}}
\providecommand{\BIBdecl}{\relax}
\BIBdecl

\bibitem{Elmqvist:2019:00}
T.~Elmqvist, E.~Andersson, N.~Frantzeskaki, T.~McPhearson, P.~Olsson,
  O.~Gaffney, K.~Takeuchi, and C.~Folke, ``Sustainability and resilience for
  transformation in the urban century,'' \emph{Nature Sustainability}, vol.~2,
  no.~4, p. 267, 2019.

\bibitem{IEA:2016:00}
{IEA}, ``{World Energy Outlook, Energy and Air Pollution},'' {International
  Energy Agency, Paris France}, Tech. Rep., 2016.

\bibitem{Birol:2013:00}
I.~E. Agency and F.~Birol, \emph{World energy outlook 2013}.\hskip 1em plus
  0.5em minus 0.4em\relax International Energy Agency Paris, 2013.

\bibitem{Commission:2011:00}
E.~Commission, ``{A Roadmap for moving to a competitive low carbon economy in
  2050},'' \emph{{European Commission, Brussel}}, 2011.

\bibitem{Rogelj:2016:00}
J.~Rogelj, M.~Den~Elzen, N.~H{\"o}hne, T.~Fransen, H.~Fekete, H.~Winkler,
  R.~Schaeffer, F.~Sha, K.~Riahi, and M.~Meinshausen, ``{Paris Agreement
  climate proposals need a boost to keep warming well below 2 C},''
  \emph{Nature}, vol. 534, no. 7609, pp. 631--639, 2016.

\bibitem{Obergassel:2016:00}
W.~Obergassel, C.~Arens, L.~Hermwille, N.~Kreibich, F.~Mersmann, H.~E. Ott, and
  H.~Wang-Helmreich, ``{Phoenix from the Ashes---An Analysis of the Paris
  Agreement to the United Nations Framework Convention on Climate Change},''
  \emph{Wuppertal Institute for Climate, Environment and Energy}, vol.~1, pp.
  1--54, 2016.

\bibitem{williams:2015:00}
J.~H. Williams, B.~Haley, F.~Kahrl, J.~Moore, A.~D. Jones, M.~S. Torn,
  H.~McJeon \emph{et~al.}, ``Pathways to deep decarbonization in the united
  states,'' 2015.

\bibitem{williams:2015:01}
J.~Williams, ``Policy implications of deep decarbonization in the united
  states,'' \emph{AGUFM}, vol. 2015, pp. PA31B--2163, 2015.

\bibitem{state-of-california:2017:00}
S.~of~California Energy~Commission \emph{et~al.}, ``California's 2030 climate
  commitment: Renewable resources for half of the state's electricity by 2030.
  state of california energy commission,'' Tech. Rep, Tech. Rep., 2017.

\bibitem{IEA:2017:00}
IEA, ``Renewables 2017 analysis and forecasts to 2022,'' {International Energy
  Agency}, Tech. Rep., October 2017.

\bibitem{LLNL:2019:00}
\BIBentryALTinterwordspacing
L.~L.~N. Laboratory. (2020) Estimated u.s energy consumption in 2019: 100.2
  quads. [Online]. Available:
  \url{https://flowcharts.llnl.gov/content/assets/images/energy/us/Energy_US_2019.png}
\BIBentrySTDinterwordspacing

\bibitem{EIA:2020:00}
EIA, ``Annual energy outlook 2020,'' \emph{Independent Statistics and Analysis,
  U.S. Energy Information Administration, Department of Energy}, 2020.

\bibitem{Kamphuis:2008:00}
R.~Kamphuis, K.~Kok, C.~Warmer, and M.~Hommelberg, ``Architectures for novel
  energy infrastructures: Multi-agent based coordination patterns,'' \emph{2008
  First International Conference on Infrastructure Systems and Services:
  Building Networks for a Brighter Future (INFRA)}, pp. 1--6, 2008.

\bibitem{hansen:2001:00}
L.~H. Hansen, P.~H. Madsen, F.~Blaabjerg, H.~Christensen, U.~Lindhard, and
  K.~Eskildsen, ``Generators and power electronics technology for wind
  turbines,'' in \emph{IECON'01. 27th Annual Conference of the IEEE Industrial
  Electronics Society (Cat. No. 37243)}, vol.~3.\hskip 1em plus 0.5em minus
  0.4em\relax IEEE, 2001, pp. 2000--2005.

\bibitem{kumar:2016:00}
Y.~Kumar, J.~Ringenberg, S.~S. Depuru, V.~K. Devabhaktuni, J.~W. Lee,
  E.~Nikolaidis, B.~Andersen, and A.~Afjeh, ``Wind energy: Trends and enabling
  technologies,'' \emph{Renewable and Sustainable Energy Reviews}, vol.~53, pp.
  209--224, 2016.

\bibitem{Munoz-Hernandez:2013:00}
G.~A. Munoz-Hernandez, S.~P. Mansoor, and D.~I. Jones, \emph{{Modelling and
  controlling hydropower plants}}.\hskip 1em plus 0.5em minus 0.4em\relax
  London: Springer, 2013.

\bibitem{Chaabene:1998:00}
M.~Chaabene and M.~Annabi, ``Dynamic thermal model for predicting solar plant
  adequate energy management,'' \emph{Energy Conversion and Management},
  vol.~39, no. 3-4, pp. 349--355, Feb 1998.

\bibitem{Pasaoglu:2012:00}
G.~Pasaoglu, M.~Honselaar, and C.~Thiel, ``Potential vehicle fleet co2
  reductions and cost implications for various vehicle technology deployment
  scenarios in europe,'' \emph{Energy Policy}, vol.~40, pp. 404--421, Jan 2012.

\bibitem{Litman:2013:00}
T.~Litman, ``Comprehensive evaluation of transport energy conservation and
  emission reduction policies,'' \emph{Transportation Research Part A: Policy
  and Practice}, vol.~47, pp. 1--23, 2013.

\bibitem{Andersen:2009:00}
P.~H. Andersen, J.~A. Mathews, and M.~Rask, ``Integrating private transport
  into renewable energy policy: The strategy of creating intelligent recharging
  grids for electric vehicles,'' \emph{Energy Policy}, vol.~37, no.~7, pp.
  2481--2486, Jul 2009.

\bibitem{Sortomme:2012:00}
E.~Sortomme and M.~A. El-Sharkawi, ``{Optimal Scheduling of Vehicle-to-Grid
  Energy and Ancillary Services},'' \emph{Smart Grid, IEEE Transactions on},
  vol.~3, no.~1, pp. 351--359, 2012.

\bibitem{U.S.-Energy-Information-Administration:2015:00}
\BIBentryALTinterwordspacing
{U.S. Energy Information Administration}, ``{Manufacturing Energy Consumption
  Survey},'' U.S. Department of Energy, Washington DC USA, Tech. Rep., 2015.
  [Online]. Available: \url{http://www.eia.gov/consumption/manufacturing/}
\BIBentrySTDinterwordspacing

\bibitem{Anair:2012:00}
D.~Anair and A.~Mahmassani, ``{State of charge: Electric vehicles' global
  warming emissions and fuel-cost savings across the United States},''
  \emph{{Union of Concerned Scientists}}, 2012.

\bibitem{Dori:2015:00}
D.~Dori, \emph{Model-based systems engineering with OPM and SysML}.\hskip 1em
  plus 0.5em minus 0.4em\relax Springer, 2015.

\bibitem{Friedenthal:2011:00}
S.~Friedenthal, A.~Moore, and R.~Steiner, \emph{{A Practical Guide to SysML:
  The Systems Modeling Language}}, 2nd~ed.\hskip 1em plus 0.5em minus
  0.4em\relax Burlington, MA: Morgan Kaufmann, 2011.

\bibitem{Cloutier:2010:00}
R.~Cloutier, G.~Muller, D.~Verma, R.~Nilchiani, E.~Hole, and M.~Bone, ``The
  concept of reference architectures,'' \emph{Systems Engineering}, vol.~13,
  no.~1, pp. 14--27, 2010.

\bibitem{Uslar:2012:00}
M.~Uslar, M.~Specht, S.~Rohjans, J.~Trefke, and J.~M. Gonz{\'a}lez, \emph{The
  Common Information Model CIM: IEC 61968/61970 and 62325-A practical
  introduction to the CIM}.\hskip 1em plus 0.5em minus 0.4em\relax Springer
  Science \& Business Media, 2012.

\bibitem{Gray:2019:00}
G.~Gray, ``Common information model primer: Fifth edition,'' EPRI, 3420
  Hillview Avenue, Palo Alto, California 94304-1338, Tech. Rep. 3002015918, May
  2019.

\bibitem{IEC:2012:00}
IEC, ``Energy management system application program interface (emsapi)---part
  301: common information model (cim) base,'' \emph{part 301: common
  information model (CIM) base}, no. 61970-301, 2012.

\bibitem{IEC:2013:00}
I.-I.~E. Commission \emph{et~al.}, ``Application integration at electric
  utilities - system interfaces for distribution management- part 11: Common
  information model (cim) extensions for distribution,'' \emph{International
  Standard}, no. IEC 61968-11, 2013.

\bibitem{IEC:2014:00}
------, ``Framework for energy market commiunications - part 301: Common
  information model (cim) extensions for markets,'' \emph{International
  Standard}, no. IEC 62325-301, 2014.

\bibitem{godfrey:2010:00}
T.~Godfrey, S.~Mullen, D.~W. Griffith, N.~Golmie, R.~C. Dugan, and C.~Rodine,
  ``Modeling smart grid applications with co-simulation,'' pp. 291--296, 2010.

\bibitem{gomes:2018:00}
C.~Gomes, C.~Thule, D.~Broman, P.~G. Larsen, and H.~Vangheluwe,
  ``Co-simulation: a survey,'' \emph{ACM Computing Surveys (CSUR)}, vol.~51,
  no.~3, pp. 1--33, 2018.

\bibitem{Rueda:2017:00}
D.~F. Rueda and E.~Calle, ``Using interdependency matrices to mitigate targeted
  attacks on interdependent networks: A case study involving a power grid and
  backbone telecommunications networks,'' \emph{International Journal of
  Critical Infrastructure Protection}, vol.~16, pp. 3--12, 2017.

\bibitem{Uday:2015:00}
P.~Uday and K.~Marais, ``Designing resilient systems-of-systems: A survey of
  metrics, methods, and challenges,'' \emph{Systems Engineering}, vol.~18,
  no.~5, pp. 491--510, 2015.

\bibitem{Thompson:2020:SPG-JR05}
D.~Thompson and A.~M. Schoonenberg, Wester C.H.and~Farid, ``{A
  Hetero-functional Graph Resilience Analysis of the Future American Electric
  Power System},'' \emph{submitted to: Nature}, vol.~1, no.~1, p.~11, 2020.

\bibitem{chassin:2005:00}
D.~P. Chassin and C.~Posse, ``Evaluating north american electric grid
  reliability using the barab{\'a}si--albert network model,'' \emph{Physica A:
  Statistical Mechanics and its Applications}, vol. 355, no. 2-4, pp. 667--677,
  2005.

\bibitem{Hernandez-Fajardo:2013:00}
I.~Hernandez-Fajardo and L.~Due{\~n}as-Osorio, ``Probabilistic study of
  cascading failures in complex interdependent lifeline systems,''
  \emph{Reliability Engineering \& System Safety}, vol. 111, pp. 260--272,
  2013.

\bibitem{ICF:2012:00}
ICF, ``Assessment of new england's natural gas pipeline capacity to satisfy
  short and near-term power generation needs,'' ICF International for ISO-NE
  Planning Advisory Committee, Tech. Rep., 2012.

\bibitem{Bonham:2020:00}
T.~Bonham, ``Resilience of unite states energy infrustructure to fluvial
  threat,'' Thesis, Dartmouth College, Hanover, NH, USA, May 2020.

\bibitem{venn:2016:00}
F.~Venn, \emph{The oil crisis}.\hskip 1em plus 0.5em minus 0.4em\relax
  Routledge, 2016.

\bibitem{action:2016:00}
P.~Action, \emph{Poor people's energy outlook 2016: National energy access
  planning from the bottom up}.\hskip 1em plus 0.5em minus 0.4em\relax
  Practical Action Publishing, 2016.

\bibitem{Gellings:2011:00}
C.~Gellings, F.~Functioning, and S.~Grid, ``Estimating the costs and benefits
  of the smart grid,'' EPRI, Palo Alto, CA, USA, Tech. Rep., 2011.

\bibitem{Gungor:2011:00}
V.~C. G{\"u}ng{\"o}r, D.~Sahin, T.~Kocak, S.~Erg{\"u}t, C.~Buccella, S.~Member,
  C.~Cecati, G.~P. Hancke, and S.~Member, ``Smart grid technologies :
  Communication technologies and standards,'' \emph{IEEE Transactions on
  Industrial Informatics}, vol.~7, no.~4, pp. 529--539, 2011.

\bibitem{Gungor:2013:00}
V.~Gungor, D.~Sahin, T.~Kocak, S.~Ergut, C.~Buccella, C.~Cecati, and G.~Hancke,
  ``A survey on smart grid potential applications and communication
  requirements,'' \emph{Industrial Informatics, IEEE Transactions on}, vol.~9,
  no.~1, pp. 28--42, Feb 2013.

\bibitem{iso-C:2012:00}
C.~ISO, ``What the duck curve tells us about managing a green grid,''
  \emph{Calif. ISO, Shap. a Renewed Futur}, pp. 1--4, 2012.

\bibitem{Joos:2000:00}
G.~Joos, B.~T. Ooi, D.~McGillis, F.~D. Galiana, and R.~Marceau, ``The potential
  of distributed generation to provide ancillary services bt - proceedings of
  the 2000 power engineering society summer meeting, july 16, 2000 - july 20,
  2000,'' ser. Proceedings of the IEEE Power Engineering Society Transmission
  and Distribution Conference, vol.~3.\hskip 1em plus 0.5em minus 0.4em\relax
  Dept. of Elec. and Comp. Engineering, Concordia University, Montreal, Que.,
  H3G-1M8, Canada: Institute of Electrical and Electronics Engineers Inc.,
  2000, pp. 1762--1767.

\bibitem{Xie:2011:00}
L.~Xie, P.~M.~S. Carvalho, L.~A. F.~M. Ferreira, J.~Liu, B.~H. Krogh, N.~Popli,
  and M.~D. Ili{{\'c}}, ``Wind integration in power systems: Operational
  challenges and possible solutions,'' \emph{Proceedings of the IEEE}, vol.~99,
  no.~1, pp. 214--232, Jan 2011.

\bibitem{Pickard:2009:00}
W.~F. Pickard, A.~Q. Shen, and N.~J. Hansing, ``Parking the power: Strategies
  and physical limitations for bulk energy storage in supply--demand matching
  on a grid whose input power is provided by intermittent sources,''
  \emph{Renewable and Sustainable Energy Reviews}, vol.~13, no.~8, pp.
  1934--1945, Oct 2009.

\bibitem{Kassakian:2011:00}
J.~Kassakian, R.~Schmalensee, G.~Desgroseilliers, T.~Heidel, K.~Afridi,
  A.~Farid, J.~Grochow, W.~Hogan, H.~Jacoby, J.~Kirtley, H.~Michaels,
  I.~Perez-Arriaga, D.~Perreault, N.~Rose, G.~Wilson, N.~Abudaldah, M.~Chen,
  P.~Donohoo, S.~Gunter, P.~Kwok, V.~Sakhrani, J.~Wang, A.~Whitaker, X.~Yap,
  R.~Zhang, and M.~I. of~Technology, ``{The Future of the Electric Grid: An
  Interdisciplinary MIT Study},'' 2011.

\bibitem{pietzcker:2017:00}
R.~C. Pietzcker, F.~Ueckerdt, S.~Carrara, H.~S. De~Boer, J.~Despr{\'e}s,
  S.~Fujimori, N.~Johnson, A.~Kitous, Y.~Scholz, P.~Sullivan \emph{et~al.},
  ``System integration of wind and solar power in integrated assessment models:
  A cross-model evaluation of new approaches,'' \emph{Energy Economics},
  vol.~64, pp. 583--599, 2017.

\bibitem{Haller:2012:00}
M.~Haller, S.~Ludig, and N.~Bauer, ``Decarbonization scenarios for the {EU} and
  {MENA} power system: Considering spatial distribution and short term dynamics
  of renewable generationnamics of renewable generation,'' \emph{{Energy
  Policy}}, vol.~47, pp. 282--290, 2012.

\bibitem{howard:2014:00}
M.~Howard, ``The integrated grid: realizing the full value of central and
  distributed energy resources,'' \emph{ICER Chron}, 2014.

\bibitem{Rogers:2014:00}
E.~a. Rogers, ``{The Energy Savings Potential of Smart Manufacturing},''
  American Council for an Energy-Efficient Economy, Washington DC USA, Tech.
  Rep. July, 2014.

\bibitem{EIA:2017:01}
\BIBentryALTinterwordspacing
EIA, ``Availability of the national energy modeling system (nems) archive,''
  United States Energy Information Administration, Tech. Rep., 2017. [Online].
  Available: \url{https://www.eia.gov/outlooks/aeo/info_nems_archive.cfm}
\BIBentrySTDinterwordspacing

\bibitem{Weilkiens:2007:00}
T.~Weilkiens, \emph{{Systems engineering with SysML/UML modeling, analysis,
  design}}.\hskip 1em plus 0.5em minus 0.4em\relax Burlington, Mass.: Morgan
  Kaufmann, 2007.

\bibitem{masters:2013:00}
G.~M. Masters, \emph{Renewable and efficient electric power systems}.\hskip 1em
  plus 0.5em minus 0.4em\relax John Wiley \& Sons, 2013.

\bibitem{mokhatab:2012:00}
S.~Mokhatab and W.~A. Poe, \emph{Handbook of natural gas transmission and
  processing}.\hskip 1em plus 0.5em minus 0.4em\relax Gulf professional
  publishing, 2012.

\bibitem{Lurie:2009:00}
M.~V. Lurie, \emph{Modeling of Oil Product and Gas Pipeline
  Transportation}.\hskip 1em plus 0.5em minus 0.4em\relax Wiley-VCH Verlag GmbH
  \& Co. KGaA, 2009.

\bibitem{EIA:2014:11}
EIA, ``Coal market module of the national energy modeling system: Model
  documentation 2014,'' \emph{Independent Statistics and Analysis, U.S. Energy
  Information Administration, Department of Energy}, 2014.

\bibitem{Albert:2004:00}
R.~Albert, I.~Albert, and G.~L. Nakarado, ``Structural vulnerability of the
  north american power grid,'' \emph{Physical review E}, vol.~69, no.~2, p.
  025103, 2004.

\bibitem{Dong:2015:00}
G.~Dong, R.~Du, L.~Tian, and R.~Liu, ``Robustness of network of networks with
  interdependent and interconnected links,'' \emph{Physica A: Statistical
  Mechanics and its Applications}, vol. 424, pp. 11--18, 2015.

\bibitem{Jean-Baptiste:2003:00}
P.~Jean-Baptiste and R.~Ducroux, ``Energy policy and climate change,''
  \emph{Energy Policy}, vol.~31, no.~2, pp. 155--166, Jan 2003.

\bibitem{Kriegler:2018:00}
E.~Kriegler, G.~Luderer, N.~Bauer, L.~Baumstark, S.~Fujimori, A.~Popp,
  J.~Rogelj, J.~Strefler, and D.~P. Van~Vuuren, ``Pathways limiting warming to
  1.5c: a tale of turning around in no time?'' \emph{Philosophical Transactions
  of the Royal Society A: Mathematical, Physical and Engineering Sciences},
  vol. 376, no. 2119, p. 20160457, 2018.

\bibitem{Mejia-Giraldo:2012:00}
\BIBentryALTinterwordspacing
D.~Mejia-Giraldo, J.~Villarreal-Marimon, Y.~Gu, Y.~He, Z.~Duan, and L.~Wang,
  ``{Sustainability and resiliency measures for long-term investment planning
  in integrated energy and transportation infrastructures},'' \emph{Journal of
  Energy Engineering}, vol. 138, no.~2, pp. 87--94, Jun. 2012. [Online].
  Available: \url{http://dx.doi.org/10.1061/(ASCE)EY.1943-7897.0000067}
\BIBentrySTDinterwordspacing

\bibitem{Robert-Lempert:2019:00}
J.~E. L. C. T. W. M. B. E. D. B.~T. Robert~Lempert, Benjamin L.~Preston,
  ``Pathways to 2050 alternative scenarios for decarbonizing the u.s.
  economy,'' Center for Climate and Energy Solutions, Climate Innovation 2050,
  2019.

\bibitem{Rogers:2013:00}
J.~Rogers, K.~Averyt, S.~Clemmer, M.~Davis, F.~Flores-Lopez, D.~Kenney,
  J.~Macknick, N.~Madden, J.~Meldrum, S.~Sattler, and E.~Spanger-Siegfried,
  ``{Water-Smart Power: Strengthening the U.S. Electricity System in a Warming
  World},'' Union for Concerned Scientists, Cambridge, MA, Tech. Rep., 2013.

\bibitem{Al-Douri:2017:00}
\BIBentryALTinterwordspacing
A.~Al-Douri, D.~Sengupta, and M.~El-Halwagi, ``Shale gas monetization - a
  review of downstream processing to chemicals and fuels,'' \emph{Journal of
  Natural Gas Science and Engineering}, 2017. [Online]. Available:
  \url{http://www.sciencedirect.com/science/article/pii/S1875510017302299}
\BIBentrySTDinterwordspacing

\bibitem{An:2003:00}
S.~An, Q.~Li, and T.~Gedra, ``{``Natural gas and electricity optimal power
  flow"},'' \emph{2003 IEEE PES Transmission and Distribution Conference and
  Exposition}, no.~l, pp. 138--143, 2003.

\bibitem{Li:2008:01}
T.~Li, M.~Eremia, M.~Shahidehpour, and M.~Shahidepour, ``Interdependency of
  natural gas network and power system security,'' \emph{IEEE Transactions on
  Power Systems}, vol.~23, no.~4, pp. 1817--1824, 2008.

\bibitem{Shahidehpour:2005:01}
M.~Shahidehpour, Y.~Fu, and T.~Wiedman, ``Impact of natural gas infrastructure
  on electric power systems,'' \emph{Proceedings of the IEEE}, vol.~93, no.~5,
  pp. 1042--1056, 2005.

\bibitem{Unsihuay:2007:00}
C.~Unsihuay, J.~W.~M. Lima, and A.~C. Z.~D. Souza, ``Modeling the integrated
  natural gas and electricity optimal power flow,'' \emph{IEEE Power
  Engineering Society General Meeting}, pp. 1--7, 2007.

\bibitem{Zlotnik:2017:00}
A.~Zlotnik, A.~Rudkevich, R.~Carter, P.~Ruiz, S.~Backhaus, and J.~Tafl, ``Grid
  architecture at the gas-electric interface,'' \emph{Los Alamos Natl. Lab.,
  Santa Fe, NM, USA, Rep. LA-UR-17-23662}, 2017.

\bibitem{Jenkins:2015:00}
S.~Jenkins, A.~Annaswamy, J.~Hansen, and J.~Knudsen, ``A dynamic model of the
  combined electricity and natural gas markets,'' in \emph{Innovative Smart
  Grid Technologies Conference (ISGT), 2015 IEEE Power \&amp; Energy
  Society}.\hskip 1em plus 0.5em minus 0.4em\relax IEEE, 2015, pp. 1--5.

\bibitem{kerr:2010:00}
R.~A. Kerr, ``Natural gas from shale bursts onto the scene,'' 2010.

\bibitem{Muzhikyan:2019:SPG-J39}
\BIBentryALTinterwordspacing
A.~Muzhikyan, S.~Muhanji, G.~Moynihan, D.~Thompson, Z.~Berzolla, and A.~M.
  Farid, ``{The 2017 ISO New England System Operational Analysis and Renewable
  Energy Integration Study},'' \emph{{Energy Reports}}, vol.~5, pp. 747--792,
  {July} 2019. [Online]. Available:
  \url{https://doi.org/10.1016/j.egyr.2019.06.005}
\BIBentrySTDinterwordspacing

\bibitem{aleklett:2010:00}
K.~Aleklett, M.~H{\"o}{\"o}k, K.~Jakobsson, M.~Lardelli, S.~Snowden, and
  B.~S{\"o}derbergh, ``The peak of the oil age--analyzing the world oil
  production reference scenario in world energy outlook 2008,'' \emph{Energy
  Policy}, vol.~38, no.~3, pp. 1398--1414, 2010.

\bibitem{AAR:2016:00}
\BIBentryALTinterwordspacing
AAR, ``Railroads and coal,'' Association of American Railroads, Tech. Rep.,
  2016. [Online]. Available:
  \url{https://www.aar.org/BackgroundPapers/Railroads%20and%20Coal.pdf}
\BIBentrySTDinterwordspacing

\bibitem{Farid:2016:EWN-J29}
\BIBentryALTinterwordspacing
A.~M. Farid, W.~N. Lubega, and W.~Hickman, ``{Opportunities for Energy-Water
  Nexus Management in the Middle East and North Africa},'' \emph{Elementa},
  vol.~4, no. 134, pp. 1--17, 2016. [Online]. Available:
  \url{http://dx.doi.org/10.12952/journal.elementa.000134}
\BIBentrySTDinterwordspacing

\bibitem{Lubega:2014:EWN-J11}
\BIBentryALTinterwordspacing
W.~N. Lubega and A.~M. Farid, ``{A Reference System Architecture for the
  Energy-Water Nexus},'' \emph{IEEE Systems Journal}, vol.~PP, no.~99, pp.
  1--11, 2014. [Online]. Available:
  \url{http://dx.doi.org/10.1109/JSYST.2014.2302031}
\BIBentrySTDinterwordspacing

\bibitem{Lubega:2014:EWN-J12}
\BIBentryALTinterwordspacing
------, ``{Quantitative Engineering Systems Model and Analysis of the
  Energy-Water Nexus},'' \emph{Applied Energy}, vol. 135, no.~1, pp. 142--157,
  2014. [Online]. Available:
  \url{http://dx.doi.org/10.1016/j.apenergy.2014.07.101}
\BIBentrySTDinterwordspacing

\bibitem{Thompson:2019:00}
J.~R. Thompson, D.~Frezza, B.~Necioglu, M.~L. Cohen, K.~Hoffman, and
  K.~Rosfjord, ``Interdependent critical infrastructure model (icim): An
  agent-based model of power and water infrastructure,'' \emph{International
  Journal of Critical Infrastructure Protection}, vol.~24, pp. 144--165, 2019.

\bibitem{Lubega:2014:EWN-T09}
W.~N. Lubega, ``{An Engineering Systems Approach to the Modeling and Analysis
  of the Energy-Water Nexus},'' Master's Thesis, Masdar Institute of Science \&
  Technology, 2014.

\bibitem{Lubega:2013:EWN-C36}
\BIBentryALTinterwordspacing
W.~N. Lubega, A.~Santhosh, A.~M. Farid, and K.~Youcef-Toumi, ``{Opportunities
  for Integrated Energy and Water Management in the GCC -- A Keynote Paper},''
  in \emph{EU-GCC Renewable Energy Policy Experts' Workshop}, no. December,
  Masdar Institute, Abu Dhabi, UAE, 2013, pp. 1--33. [Online]. Available:
  \url{http://www.grc.net/data/contents/uploads/Opportunities\_for\_Integrated\_Energy\_and\_Water\_Management\_in\_the\_GCC\_5874.pdf}
\BIBentrySTDinterwordspacing

\bibitem{Lubega:2014:EWN-C35}
\BIBentryALTinterwordspacing
------, ``{An Integrated Energy and Water Market for the Supply Side of the
  Energy-Water Nexus in the Engineered Infrastructure},'' in \emph{ASME 2014
  Power Conference}, Baltimore, MD, 2014, pp. 1--11. [Online]. Available:
  \url{http://dx.doi.org/10.1115/POWER2014-32075}
\BIBentrySTDinterwordspacing

\bibitem{Lubega:2014:EWN-C34}
\BIBentryALTinterwordspacing
W.~N. Lubega and A.~M. Farid, ``{An Engineering Systems Sensitivity Analysis
  Model for Holistic Energy-Water Nexus Planning},'' in \emph{ASME 2014 Power
  Conference}, Baltimore, MD, 2014, pp. 1--10. [Online]. Available:
  \url{http://dx.doi.org/10.1115/POWER2014-32076}
\BIBentrySTDinterwordspacing

\bibitem{Farid:2013:EWN-C15}
\BIBentryALTinterwordspacing
A.~M. Farid and W.~N. Lubega, ``{Powering and Watering Agriculture: Application
  of Energy-Water Nexus Planning},'' in \emph{GHTC 2013: IEEE Global
  Humanitarian Technology Conference}, Silicon Valley, CA, USA, 2013, pp. 1--6.
  [Online]. Available:
  \url{http://dx.doi.org.libproxy.mit.edu/10.1109/GHTC.2013.6713689}
\BIBentrySTDinterwordspacing

\bibitem{Santhosh:2012:EWN-C09}
\BIBentryALTinterwordspacing
A.~Santhosh, A.~M. Farid, A.~Adegbege, and K.~Youcef-Toumi, ``{Simultaneous
  Co-optimization for the Economic Dispatch of Power and Water Networks},'' in
  \emph{The 9th IET International Conference on Advances in Power System
  Control, Operation and Management}, Hong Kong, China, 2012, pp. 1--6.
  [Online]. Available: \url{http://dx.doi.org/10.1049/cp.2012.2148}
\BIBentrySTDinterwordspacing

\bibitem{Santhosh:2013:EWN-C26}
\BIBentryALTinterwordspacing
A.~Santhosh, A.~M. Farid, and K.~Youcef-Toumi, ``{The Impact of Storage
  Facilities on the Simultaneous Economic Dispatch of Power and Water Networks
  Limited by Ramping Rates},'' in \emph{IEEE International Conference on
  Industrial Technology}, Cape Town, South Africa, 2013, pp. 1--6. [Online].
  Available: \url{http://dx.doi.org/10.1109/ICIT.2013.6505794}
\BIBentrySTDinterwordspacing

\bibitem{Santhosh:2014:EWN-J10}
\BIBentryALTinterwordspacing
------, ``{The Impact of Storage Facility Capacity and Ramping Capabilities on
  the Supply Side of the Energy-Water Nexus},'' \emph{Energy}, vol.~66, no.~1,
  pp. 1--10, 2014. [Online]. Available:
  \url{http://dx.doi.org/10.1016/j.energy.2014.01.031}
\BIBentrySTDinterwordspacing

\bibitem{Hickman:2017:EWN-J32}
\BIBentryALTinterwordspacing
W.~Hickman, A.~Muzhikyan, and A.~M. Farid, ``{The Synergistic Role of Renewable
  Energy Integration into the Unit Commitment of the Energy Water Nexus},''
  \emph{Renewable Energy}, vol. 108, no.~1, pp. 220--229, 2017. [Online].
  Available: \url{https://dx.doi.org/10.1016/j.renene.2017.02.063}
\BIBentrySTDinterwordspacing

\bibitem{Murrant:2015:00}
D.~Murrant, A.~Quinn, and L.~Chapman, ``The water-energy nexus: future water
  resource availability and its implications on uk thermal power generation,''
  \emph{Water and Environment Journal}, 2015.

\bibitem{Wakeel:2016:00}
\BIBentryALTinterwordspacing
M.~Wakeel and B.~Chen, ``Energy consumption in urban water cycle,''
  \emph{Energy Procedia}, vol. 104, pp. 123 -- 128, 2016. [Online]. Available:
  \url{http://www.sciencedirect.com/science/article/pii/S1876610216315776}
\BIBentrySTDinterwordspacing

\bibitem{Macknick:2012:00}
J.~Macknick, R.~Newmark, G.~Heath, and K.~C. Hallett, ``Operational water
  consumption and withdrawal factors for electricity generating technologies: a
  review of existing literature,'' \emph{Environmental Research Letters},
  vol.~7, no.~4, p. 045802, Dec 2012.

\bibitem{Macknick:2012:01}
J.~Macknick, S.~Sattler, K.~Averyt, S.~Clemmer, and J.~Rogers, ``The water
  implications of generating electricity: water use across the united states
  based on different electricity pathways through 2050,'' \emph{Environmental
  Research Letters}, vol.~7, no.~4, p. 045803, Dec 2012.

\bibitem{Averyt:2011:00}
K.~Averyt, J.~Fisher, A.~Huber-Lee, A.~Lewis, J.~Macknick, N.~Madden,
  J.~Rogers, and S.~Tellinghuisen, ``Freshwater use by us power plants:
  Electricity's thirst for a precious resource,'' Union of Concerned
  Scientists, Cambridge, MA, USA, Tech. Rep., 2011.

\bibitem{Tidwell:2014:00}
\BIBentryALTinterwordspacing
V.~C. Tidwell, J.~Macknick, K.~Zemlick, J.~Sanchez, and T.~Woldeyesus,
  ``{Transitioning to zero freshwater withdrawal in the U.S. for thermoelectric
  generation},'' \emph{Applied Energy}, vol. 131, pp. 508--516, Jun. 2014.
  [Online]. Available:
  \url{http://linkinghub.elsevier.com/retrieve/pii/S0306261913009215}
\BIBentrySTDinterwordspacing

\bibitem{Roosa:2008:00}
S.~A. Roosa, \emph{Sustainable Development Handbook}.\hskip 1em plus 0.5em
  minus 0.4em\relax 700 Indian Trail Lilbum, GA: The Fairmont Press, Inc.,
  2008.

\bibitem{mancarella:2016:00}
P.~Mancarella, G.~Andersson, J.~Pe{\c{c}}as-Lopes, and K.~R. Bell, ``Modelling
  of integrated multi-energy systems: Drivers, requirements, and
  opportunities,'' in \emph{2016 Power Systems Computation Conference
  (PSCC)}.\hskip 1em plus 0.5em minus 0.4em\relax IEEE, 2016, pp. 1--22.

\bibitem{subramanian:2018:00}
A.~S.~R. Subramanian, T.~Gundersen, and T.~A. Adams, ``Modeling and simulation
  of energy systems: A review,'' \emph{Processes}, vol.~6, no.~12, p. 238,
  2018.

\bibitem{Beuzekom:2015:00}
I.~{van Beuzekom}, M.~{Gibescu}, and J.~G. {Slootweg}, ``A review of
  multi-energy system planning and optimization tools for sustainable urban
  development,'' in \emph{2015 IEEE Eindhoven PowerTech}, 2015, pp. 1--7.

\bibitem{Geidl:2007:00}
M.~Geidl and G.~Andersson, ``Optimal power flow of multiple energy carriers,''
  \emph{IEEE Transactions on Power Systems}, vol.~22, no.~1, pp. 145--155,
  2007.

\bibitem{Krause:2011:00}
T.~Krause, G.~Andersson, K.~Fr{\"o}hlich, and A.~Vaccaro, ``Multiple-energy
  carriers: Modeling of production, delivery, and consumption,''
  \emph{Proceedings of the IEEE}, vol.~99, no.~1, pp. 15--27, 2011.

\bibitem{Ma:2018:00}
\BIBentryALTinterwordspacing
T.~Ma, J.~Wu, L.~Hao, W.-J. Lee, H.~Yan, and D.~Li, ``The optimal structure
  planning and energy management strategies of smart multi energy systems,''
  \emph{Energy}, vol. 160, pp. 122--141, 2018. [Online]. Available:
  \url{https://www.sciencedirect.com/science/article/pii/S0360544218312635}
\BIBentrySTDinterwordspacing

\bibitem{MANCARELLA:2014:00}
\BIBentryALTinterwordspacing
P.~Mancarella, ``Mes (multi-energy systems): An overview of concepts and
  evaluation models,'' \emph{Energy}, vol.~65, pp. 1--17, 2014. [Online].
  Available:
  \url{https://www.sciencedirect.com/science/article/pii/S0360544213008931}
\BIBentrySTDinterwordspacing

\bibitem{Quelhas:2007:00}
A.~Quelhas, E.~Gil, J.~D. McCalley, and S.~M. Ryan, ``A multiperiod generalized
  network flow model of the u.s. integrated energy system: Part i---model
  description,'' \emph{IEEE Transactions on Power Systems}, vol.~22, no.~2, pp.
  829--836, 2007.

\bibitem{WU:2016:00}
\BIBentryALTinterwordspacing
J.~Wu, J.~Yan, H.~Jia, N.~Hatziargyriou, N.~Djilali, and H.~Sun, ``Integrated
  energy systems,'' \emph{Applied Energy}, vol. 167, pp. 155--157, 2016.
  [Online]. Available:
  \url{https://www.sciencedirect.com/science/article/pii/S0306261916302124}
\BIBentrySTDinterwordspacing

\bibitem{jacobson:2009:00}
M.~Z. Jacobson, ``Review of solutions to global warming, air pollution, and
  energy security,'' \emph{Energy \& Environmental Science}, vol.~2, no.~2, pp.
  148--173, 2009.

\bibitem{jacobson:2009:01}
M.~Z. Jacobson and M.~A. Delucchi, ``A path to sustainable energy by 2030,''
  \emph{Scientific American}, vol. 301, no.~5, pp. 58--65, 2009.

\bibitem{jacobson:2015:00}
M.~Z. Jacobson, M.~A. Delucchi, G.~Bazouin, Z.~A. Bauer, C.~C. Heavey,
  E.~Fisher, S.~B. Morris, D.~J. Piekutowski, T.~A. Vencill, and T.~W. Yeskoo,
  ``100\% clean and renewable wind, water, and sunlight (wws) all-sector energy
  roadmaps for the 50 united states,'' \emph{Energy \& Environmental Science},
  vol.~8, no.~7, pp. 2093--2117, 2015.

\bibitem{jenkins:2017:00}
J.~D. Jenkins and N.~A. Sepulveda, ``Enhanced decision support for a changing
  electricity landscape: the genx configurable electricity resource capacity
  expansion model,'' \emph{An MIT Energy Initiative Working Paper.
  https://energy. mit.
  edu/wpcontent/uploads/2017/10/Enhanced-Decision-Support-for-a-Changing-Electricity-Landscape.
  pdf}, 2017.

\bibitem{jenkins:2018:00}
J.~D. Jenkins, M.~Luke, and S.~Thernstrom, ``Getting to zero carbon emissions
  in the electric power sector,'' \emph{Joule}, vol.~2, no.~12, pp. 2498--2510,
  2018.

\bibitem{Lubega:2013:EWN-C19}
\BIBentryALTinterwordspacing
W.~N. Lubega and A.~M. Farid, ``{A Meta-System Architecture for the
  Energy-Water Nexus},'' in \emph{8th Annual IEEE Systems of Systems
  Conference}, Maui, Hawaii, USA, 2013, pp. 1--6. [Online]. Available:
  \url{http://dx.doi.org/10.1109/SYSoSE.2013.6575246}
\BIBentrySTDinterwordspacing

\bibitem{Abdulla:2015:EWN-C53}
\BIBentryALTinterwordspacing
H.~Abdulla and A.~M. Farid, ``Extending the energy-water nexus reference
  architecture to the sustainable development of agriculture, industry \&
  commerce,'' in \emph{First IEEE International Smart Cities Conference},
  Guadalajara, Mexico, 2015, pp. 1--7. [Online]. Available:
  \url{http://dx.doi.org/10.1109/ISC2.2015.7366166}
\BIBentrySTDinterwordspacing

\bibitem{glasson:2013:00}
J.~Glasson and R.~Therivel, \emph{Introduction to environmental impact
  assessment}.\hskip 1em plus 0.5em minus 0.4em\relax Routledge, 2013.

\bibitem{Platts:2017:00}
\BIBentryALTinterwordspacing
Platts, ``Platts energy map data pro,'' {S\&P Global Platts}, Tech. Rep., 2017.
  [Online]. Available:
  \url{https://www.spglobal.com/platts/en/products-services/oil/map-data-pro}
\BIBentrySTDinterwordspacing

\bibitem{Reichwein:2012:00}
\BIBentryALTinterwordspacing
A.~Reichwein and C.~Paredis, ``Magicdraw sysml-modelica integration: Java-based
  implementation of the omg sysml-modelica transformation (sym) using magicdraw
  sysml,'' Object Management Group, Tech. Rep., 2012. [Online]. Available:
  \url{http://www.mbsec.gatech.edu/research/projects/active/sysml-modelica-integration}
\BIBentrySTDinterwordspacing

\bibitem{SE-Handbook-Working-Group:2011:00}
{SE Handbook Working Group}, \emph{Systems Engineering Handbook: A Guide for
  System Life Cycle Processes and Activities}.\hskip 1em plus 0.5em minus
  0.4em\relax International Council on Systems Engineering (INCOSE), 2011.

\bibitem{Crawley:2015:00}
E.~Crawley, B.~Cameron, and D.~Selva, \emph{System Architecture: Strategy and
  Product Development for Complex Systems}.\hskip 1em plus 0.5em minus
  0.4em\relax Upper Saddle River, N.J.: Prentice Hall Press, 2015.

\bibitem{Rumbaugh:2005:00}
J.~Rumbaugh, I.~Jacobson, and G.~Booch, \emph{{The Unified Modeling Language
  Reference Manual}}.\hskip 1em plus 0.5em minus 0.4em\relax Reading, Mass.:
  Addison-Wesley, 2005.

\bibitem{rosa:2017:00}
R.~N. Rosa, ``The role of synthetic fuels for a carbon neutral economy,''
  \emph{C---Journal of Carbon Research}, vol.~3, no.~2, p.~11, 2017.

\bibitem{De-Weck:2011:00}
\BIBentryALTinterwordspacing
O.~L. {De Weck}, D.~Roos, and C.~L. Magee, \emph{{Engineering systems: meeting
  human needs in a complex technological world}}.\hskip 1em plus 0.5em minus
  0.4em\relax Cambridge, Mass.: MIT Press, 2011. [Online]. Available:
  \url{http://www.knovel.com/knovel2/Toc.jsp?BookID=4611
  http://mitpress-ebooks.mit.edu/product/engineering-systems}
\BIBentrySTDinterwordspacing

\bibitem{Rinaldi:2004:00}
S.~M. Rinaldi, ``Modeling and simulating critical infrastructures and their
  interdependencies,'' in \emph{System Sciences, 2004. Proceedings of the 37th
  Annual Hawaii International Conference on}, {Jan} 2004, pp. 8 pp.--.

\bibitem{Prasad:2014:00}
R.~D. Prasad, R.~Bansal, and A.~Raturi, ``Multi-faceted energy planning: A
  review,'' \emph{Renewable and Sustainable Energy Reviews}, vol.~38, pp.
  686--699, 2014.

\bibitem{buldyrev:2010:00}
S.~V. Buldyrev, R.~Parshani, G.~Paul, H.~E. Stanley, and S.~Havlin,
  ``Catastrophic cascade of failures in interdependent networks,''
  \emph{Nature}, vol. 464, no. 7291, p. 1025, 2010.

\bibitem{unruh:2000:00}
G.~C. Unruh, ``Understanding carbon lock-in,'' \emph{Energy policy}, vol.~28,
  no.~12, pp. 817--830, 2000.

\bibitem{unruh:2002:00}
------, ``Escaping carbon lock-in,'' \emph{Energy policy}, vol.~30, no.~4, pp.
  317--325, 2002.

\bibitem{seto:2016:00}
K.~C. Seto, S.~J. Davis, R.~B. Mitchell, E.~C. Stokes, G.~Unruh, and
  D.~{\"U}rge-Vorsatz, ``Carbon lock-in: types, causes, and policy
  implications,'' \emph{Annual Review of Environment and Resources}, vol.~41,
  2016.

\bibitem{markolf:2018:00}
S.~A. Markolf, M.~V. Chester, D.~A. Eisenberg, D.~M. Iwaniec, C.~I. Davidson,
  R.~Zimmerman, T.~R. Miller, B.~L. Ruddell, and H.~Chang, ``Interdependent
  infrastructure as linked social, ecological, and technological systems
  (setss) to address lock-in and enhance resilience,'' \emph{Earth's Future},
  vol.~6, no.~12, pp. 1638--1659, 2018.

\bibitem{Palensky:2011:00}
P.~Palensky and D.~Dietrich, ``{Demand Side Management: Demand Response,
  Intelligent Energy Systems, and Smart Loads},'' \emph{Industrial Informatics,
  IEEE Transactions on}, vol.~7, no.~3, pp. 381--388, 2011.

\bibitem{Xie:2009:01}
L.~Xie, M.~D. Ilic, and M.~D. Ili, ``{Model Predictive Economic / Environmental
  Dispatch of Power Systems with Intermittent Resources},'' in \emph{2009 Power
  \& Energy Society General Meeting}, 2009, pp. 1--6.

\bibitem{alabdulwahab:2015:00}
A.~Alabdulwahab, A.~Abusorrah, X.~Zhang, and M.~Shahidehpour, ``Coordination of
  interdependent natural gas and electricity infrastructures for firming the
  variability of wind energy in stochastic day-ahead scheduling,'' \emph{IEEE
  Transactions on Sustainable Energy}, vol.~6, no.~2, pp. 606--615, 2015.

\bibitem{MACKINNON:2018:00}
\BIBentryALTinterwordspacing
M.~A. {Mac Kinnon}, J.~Brouwer, and S.~Samuelsen, ``The role of natural gas and
  its infrastructure in mitigating greenhouse gas emissions, improving regional
  air quality, and renewable resource integration,'' \emph{Progress in Energy
  and Combustion Science}, vol.~64, pp. 62 -- 92, 2018. [Online]. Available:
  \url{http://www.sciencedirect.com/science/article/pii/S0360128517300680}
\BIBentrySTDinterwordspacing

\bibitem{Lubega:2013:EWN-BC02}
\BIBentryALTinterwordspacing
W.~N. Lubega and A.~M. Farid, ``{An engineering systems model for the
  quantitative analysis of the energy-water nexus},'' in \emph{Complex Systems
  Design \& Management}.\hskip 1em plus 0.5em minus 0.4em\relax Paris, France:
  Springer Berlin Heidelberg, 2013, ch.~16, pp. 219--231. [Online]. Available:
  \url{http://dx.doi.org/10.1007/978-3-319-02812-5_16}
\BIBentrySTDinterwordspacing

\bibitem{EIA:2019:00}
EIA, ``The national energy modeling system: An overview 2018,''
  \emph{Independent Statistics and Analysis, U.S. Energy Information
  Administration, Department of Energy}, 2019.

\bibitem{Schoonenberg:2018:ISC-BK04}
\BIBentryALTinterwordspacing
W.~C. Schoonenberg, I.~S. Khayal, and A.~M. Farid, \emph{{A Hetero-functional
  Graph Theory for Modeling Interdependent Smart City Infrastructure}}.\hskip
  1em plus 0.5em minus 0.4em\relax Berlin, Heidelberg: Springer, 2018.
  [Online]. Available: \url{http://dx.doi.org/10.1007/978-3-319-99301-0}
\BIBentrySTDinterwordspacing

\bibitem{schoonenberg:2021:00}
W.~C.~H. Schoonenberg and A.~M. Farid, ``Hetero-functional network minimum cost
  flow optimization: A hydrogen-natural gas network example,'' 2021.

\bibitem{Milano:2010:17}
\BIBentryALTinterwordspacing
F.~Milano, \emph{{Power system modelling and scripting}}, 1st~ed.\hskip 1em
  plus 0.5em minus 0.4em\relax New York: Springer, 2010. [Online]. Available:
  \url{http://www.uclm.es/area/gsee/web/Federico/psat.htm}
\BIBentrySTDinterwordspacing

\bibitem{Thompson:2020:SPG-C68}
\BIBentryALTinterwordspacing
D.~Thompson, W.~C. Schoonenberg, and A.~M. Farid, ``{A Hetero-functional Graph
  Analysis of Electric Power System Structural Resilience},'' in \emph{IEEE
  Innovative Smart Grid Technologies Conference North America}, Washington, DC,
  United states, 2020, pp. 1--5. [Online]. Available:
  \url{http://dx.doi.org/10.1109/ISGT45199.2020.9087732}
\BIBentrySTDinterwordspacing

\end{thebibliography}

\begin{IEEEbiography}[{\includegraphics[width=1in,clip,keepaspectratio]{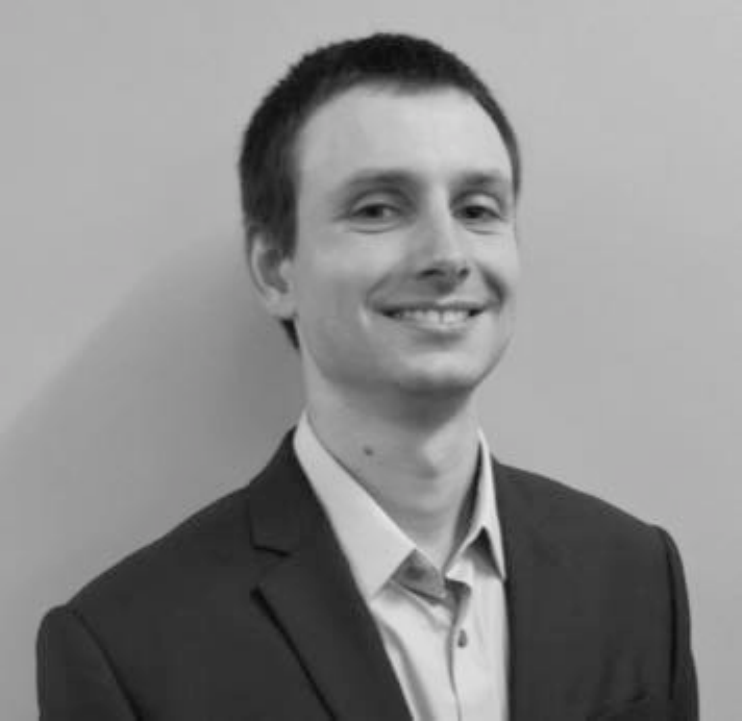}}]{Dakota J. Thompson} graduated from Colby College in 2018 with a B.A.  in Physics and minor in Computer Science. He is now pursuing a Ph.D. in energy systems engineering at the Thayer School of Engineering at Dartmouth. As an undergraduate, Dakota worked on several reseearch projects with the LIINES at the Thayer School of Engineering at Dartmouth and continues his research in power grid resilience, renewable energy integration, and hetero-functional graph theory.
\end{IEEEbiography}

\begin{IEEEbiography}[{\includegraphics[width=1in,height=1.25in,clip,keepaspectratio]{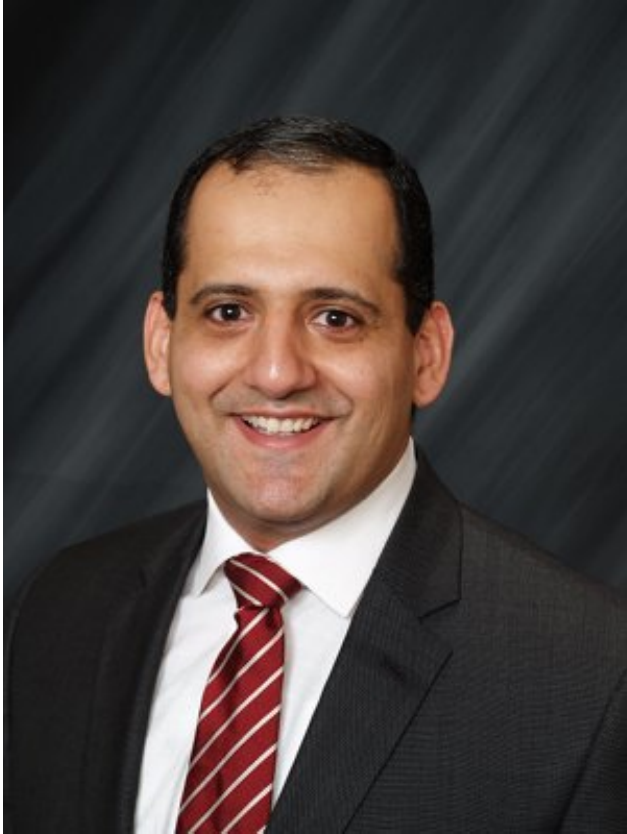}}]{Amro M. Farid} 
is currently an Associate Professor of Engineering at the Thayer School of Engineering at Dartmouth and Adjunct Associate Professor of computer science at the Department of Computer Science. He leads the Laboratory for Intelligent Integrated Networks of Engineering Systems  (LIINES). The laboratory maintains an active research program in Smart Power Grids, Energy-Water Nexus, Energy-Transportation Nexus, Industrial Energy Management,  and Integrated Smart City Infrastructures. He received his Sc. B. in 2000 and his Sc. M. 2002 in mechanical engineering from MIT and his Ph.D. degree in Engineering from the U. of Cambridge (UK).
\vspace{-0.1in}
\end{IEEEbiography}

\end{document}